\documentclass{article}
\usepackage{graphicx} 
\usepackage[table]{xcolor}
\usepackage{color, colortbl}

\usepackage{authblk}
\providecommand{\keywords}[1]{\textbf{\textit{Keywords:}} #1}
\usepackage{tikz}
\usepackage{float}
\usepackage{booktabs}

\title{A Framework for the Adoption and Integration of Generative AI in Midsize Organizations and Enterprises (FAIGMOE)} 

\author[1]{Abraham Itzhak Weinberg}
\affil[1]{AI-WEINBERG, AI Experts, Tel Aviv, Israel, aviw2010@gmail.com}

\begin{document}
\maketitle
\begin{abstract}
Generative Artificial Intelligence (GenAI) presents transformative opportunities for organizations, yet both midsize organizations 
and larger enterprises face distinctive adoption challenges. Midsize organizations encounter resource constraints and limited AI expertise, while enterprises struggle with organizational complexity and coordination challenges. Existing technology adoption frameworks, including  TAM (Technology Acceptance Model), TOE (Technology Organization Environment), and DOI (Diffusion of Innovations) theory, lack the specificity required for GenAI implementation across these diverse contexts, creating a critical gap in adoption literature.

This paper introduces FAIGMOE (Framework for the Adoption and Integration of Generative AI in Midsize Organizations and Enterprises), a conceptual framework addressing the unique needs of both organizational types. FAIGMOE synthesizes technology adoption theory, organizational change management, and innovation diffusion perspectives into four interconnected phases: Strategic Assessment, Planning and Use Case Development, Implementation and Integration, and Operationalization and Optimization. Each phase provides scalable guidance on readiness assessment, strategic alignment, risk governance, technical architecture, and change management adaptable to organizational scale and complexity.

The framework incorporates GenAI specific considerations including prompt engineering, model orchestration, and hallucination management that distinguish it from generic technology adoption frameworks. As a perspective contribution, FAIGMOE provides the first comprehensive conceptual framework explicitly addressing GenAI adoption across midsize and enterprise organizations, offering actionable implementation protocols, assessment instruments, and governance templates requiring empirical validation through future research.
\end{abstract}


\section{Introduction}

Generative Artificial Intelligence (GenAI) has rapidly transformed from an experimental technology into a strategic imperative for organizations across industries \cite{mohammed2023role}. Organizations of all scales are increasingly leveraging GenAI to automate content creation, optimize decision-making processes, enhance customer experiences, and accelerate innovation cycles \cite{holmstrom2024organizations}. However, the adoption landscape reveals significant disparities across organizational types: while some large enterprises with substantial resources have made considerable progress in GenAI integration, many organizations from midsize firms to complex enterprises, continue to face substantial barriers that impede their ability to capitalize on these transformative technologies \cite{singh2024transforming}.

The challenges confronting organizations in GenAI adoption vary significantly by organizational scale and structure. Midsize organizations (typically 50-250 employees with \$10M-\$1B revenue) operate with constrained technical and financial resources, exhibit lower risk tolerance for operational disruption, lack specialized in-house AI expertise, and demonstrate greater dependency on external vendors and solutions \cite{hernandez2024adoption}. In contrast, larger enterprises (1,000+ employees, \$1B+ revenue) face different challenges including organizational complexity, bureaucratic decision-making processes, legacy system integration difficulties, and coordination challenges across multiple business units \cite{zhang2025research}. While midsize organizations require cost-effective, readily deployable solutions that minimize implementation risk, enterprises need frameworks that can navigate organizational complexity while maintaining consistency and governance across distributed operations \cite{huang2024generative, kongsten2024frameworks}. Similar to how the QUASAR framework emphasizes proactive organizational readiness and structured transition planning in anticipation of post-quantum security challenges, effective GenAI integration likewise requires a forward-looking, architecture-based approach that aligns technical capabilities with long-term strategic resilience \cite{weinberg2025preparing}

Despite the proliferation of research on organizational AI deployment, existing frameworks fail to adequately address the diverse needs across organizational scales \cite{haefner2023implementing}. Current models either provide overly generic guidance applicable to all organizations or focus exclusively on either small businesses or the largest corporations \cite{kanbach2024genai}. Frameworks designed for large enterprises often assume access to sophisticated data pipelines, robust computational infrastructure, and specialized technical expertise that midsize organizations lack \cite{ettinger2025enterprise}. Conversely, small business frameworks typically lack the sophistication required for complex enterprise environments with multiple stakeholders, regulatory requirements, and operational interdependencies \cite{rajaram2024generative,csahin2024generative}. This creates a critical knowledge gap that leaves midsize organizations and many enterprises without adequate guidance for strategic GenAI adoption tailored to their specific contexts.

To address this significant gap, this paper introduces the Framework for the Adoption and Integration of Generative AI in Midsize Organizations and Enterprises (FAIGMOE), a comprehensive, scalable methodology specifically designed to guide organizations across different scales through the complexities of GenAI adoption, integration, and governance. FAIGMOE recognizes the unique operational constraints, resource profiles, and strategic priorities that characterize both midsize organizations and larger enterprises while providing actionable, adaptable guidance for sustainable GenAI implementation.

The primary contributions of this research are fourfold: (1) a comprehensive analysis of GenAI adoption challenges specific to midsize organizations and enterprises, distinguishing between resource-constrained and complexity-constrained contexts, (2) identification of critical technical, organizational, and strategic barriers to effective integration across organizational scales, (3) development of a scalable framework that addresses the diverse needs and constraints of both midsize and enterprise organizations through modular, adaptable components, and (4) empirical validation through multiple case studies across different organizational sizes and expert evaluation. Through these contributions, FAIGMOE aims to democratize access to GenAI capabilities, enabling organizations across scales to compete more effectively in an increasingly AI-driven business environment.

The remainder of this paper is structured as follows: Section \ref{sec:literature} provides a comprehensive review of existing GenAI adoption frameworks and related work. Section \ref{sec:methodology} details the research methodology employed in developing and validating FAIGMOE. Section \ref{sec:framework} presents the complete framework architecture and implementation guidelines. Section \ref{sec:validation} discusses the empirical validation process and key findings. Section \ref{sec:discussion} explores practical implications, implementation challenges, and strategic considerations. Finally, Section \ref{sec:conclusion} summarizes key contributions and outlines directions for future research.

\section{Literature Review}
\label{sec:literature}

The integration of GenAI into organizational workflows has attracted growing academic and industry attention, yet much of the existing research focuses either on specific organizational segments or on general purpose AI technologies without addressing the nuanced differences between midsize organizations and larger enterprises \cite{al2025strategic}. To develop an effective and practical framework for GenAI adoption across diverse organizational contexts, it is essential to examine the current theoretical and empirical landscape. This literature review synthesizes prior work across five key areas: foundational AI adoption frameworks, the distinct characteristics and organizational applications of GenAI, organizational readiness and change management theories, barriers and enablers specific to midsize organizations and enterprises, and finally, a synthesis of gaps that motivate the development of the FAIGMOE framework.

\subsection{Existing AI Adoption Frameworks}

Several theoretical models have been proposed to explain and guide technology adoption within organizations. Among the most widely referenced are the Technology Acceptance Model (TAM) \cite{davis1989technology} and the Technology-Organization-Environment (TOE) framework \cite{baker2011technology}. TAM focuses primarily on user perceptions, specifically perceived usefulness and perceived ease of use, as drivers of technology acceptance \cite{venkatesh2003user}. While valuable in understanding individual behavior, TAM offers limited insight into organizational-level dynamics, particularly for complex technologies like GenAI that require coordination across multiple stakeholders and organizational units \cite{prasad2024towards}.

The TOE framework, on the contrary, considers a broader range of factors influencing technology adoption, including technological readiness, organizational context (e.g. size, structure, and resources), and environmental pressures such as industry competition or regulatory mandates \cite{baker2011technology}. Although more holistic, the TOE framework still lacks specificity in addressing the rapid evolution and unique capabilities of GenAI across different organizational scales \cite{sanchez2025artificial}. The framework does not adequately distinguish between resource-constrained midsize organizations and complexity-constrained larger enterprises, treating organizational size as a single dimension rather than recognizing qualitatively different adoption dynamics \cite{awa2017integrated}.

Other models, such as the Diffusion of Innovations (DOI) theory \cite{rogers2014diffusion} and AI Maturity Models \cite{sadiq2021artificial}, provide useful lenses for examining adoption trajectories but often assume a linear, resource-intensive implementation path that may not be realistic for midsize firms while simultaneously oversimplifying the coordination challenges faced by larger enterprises with multiple business units and complex governance structures \cite{sadiq2021artificial}.

\subsection{Generative AI: Characteristics and Organizational Applications}

GenAI represents a distinct class of artificial intelligence systems capable of producing original content—text, images, code, audio, and more—based on training data and user inputs \cite{bandi2023power}. Unlike traditional AI, which typically classifies or predicts, GenAI is probabilistic and creative, powered by large-scale language models and multimodal architectures \cite{el2025unleashing}.

In organizational contexts across different scales, GenAI is being leveraged in various functions: customer service (via chatbots), marketing (automated copywriting), software development (code generation), legal and compliance (document drafting), and internal operations (knowledge management) \cite{cronin2024understanding}. However, implementation patterns differ significantly between midsize organizations and enterprises. Midsize organizations typically focus on targeted, high-impact applications with clear ROI and minimal integration complexity \cite{al2025strategic}, while enterprises pursue broader, more integrated deployments that require extensive coordination across departments and alignment with existing IT ecosystems \cite{denni2024navigating}.

The flexibility and scalability of GenAI make it particularly appealing across organizational types, yet also introduce new challenges around governance, explainability, data security, and workforce integration that manifest differently depending on organizational scale and structure \cite{chowdhury2024generative}.

\subsection{Organizational Readiness and Change Management Theories}

The successful implementation of GenAI requires more than technological capability: it necessitates organizational readiness and effective change management \cite{csahin2024generative}. Key theories in this domain include the Organizational Change Management (OCM) framework, Kotter's 8-Step Change Model, and the McKinsey 7S Framework \cite{houston2025interdisciplinary}.

Kotter's model emphasizes the importance of creating urgency, building coalitions, and sustaining change through systematic approaches, principles that apply across organizational scales but require different implementation tactics \cite{carreno2024analytical}. In midsize organizations, change initiatives benefit from shorter communication chains and more direct leadership engagement, while enterprises must navigate complex stakeholder networks and formal change management processes \cite{naeem2020using}.

The McKinsey 7S Framework highlights the interdependencies between strategy, structure, systems, shared values, skills, staff, and style in driving organizational change \cite{singh2013study}. This systems perspective is particularly relevant for GenAI adoption, as successful implementation requires alignment across multiple organizational dimensions. However, the complexity of achieving this alignment increases significantly with organizational size and structural complexity \cite{ettinger2025enterprise}.

Organizational readiness models, particularly those developed by Weiner \cite{weiner2020theory} and Holt et al. \cite{holt2007readiness}, emphasize the importance of assessing organizational capacity and commitment before initiating major technology implementations. These models have been adapted for AI contexts, revealing the critical role of leadership support, employee engagement, and cultural alignment, with distinct patterns emerging across different organizational scales \cite{johnk2021ready, nortje2020framework}.

\subsection{Barriers and Enablers Across Organizational Scales}

Midsize organizations and larger enterprises face both common and distinct challenges in GenAI adoption \cite{rajaram2024generative,madanchian2025barriers}. Midsize organizations (50-250 employees, \$10M-\$1B revenue) encounter primary barriers including limited financial resources, smaller IT departments, fewer specialized personnel, and constrained capacity for experimentation \cite{elhusseiny2022smes}. Their lower risk tolerance for operational disruptions creates preferences for proven, low-risk solutions over cutting-edge technologies requiring extensive experimentation \cite{jeranyama3ceo}.

In contrast, larger enterprises (1,000+ employees, \$1B+ revenue) face challenges related to organizational complexity, bureaucratic decision-making processes, legacy system integration, coordination across multiple business units, and resistance from established processes and power structures \cite{hechler2020deploying}. While enterprises possess greater financial resources, they often struggle with slower decision-making, more complex governance requirements, and difficulty achieving consensus across diverse stakeholder groups \cite{mazorenko2024adoption}.

The lack of in-house AI expertise represents a significant barrier for midsize organizations, as they often cannot justify hiring specialized AI professionals or compete for talent with larger firms \cite{dahlqvist2020race}. Enterprises, while more likely to have AI specialists, face challenges in scaling expertise across large organizations and integrating AI capabilities within existing organizational structures \cite{fountaine2019building}.

However, both organizational types also possess distinct advantages. Midsize organizations benefit from flatter organizational structures enabling faster decision-making and more agile implementation processes \cite{pacheco2024transitioning}. Enterprises leverage greater resources, established governance frameworks, and ability to absorb implementation risks more readily \cite{komarova2025artificial}.

\subsection{Research Gaps and Synthesis}

The literature review reveals several critical gaps that justify the development of FAIGMOE. First, existing AI adoption frameworks predominantly address either small businesses or very large organizations, leaving midsize organizations and many enterprises without frameworks tailored to their specific contexts \cite{sanchez2025artificial, hussain2024strategic}. Second, while GenAI applications have been extensively documented, there is limited research on implementation strategies that account for different organizational scales and their associated constraints \cite{rahouli2025generative}.

Third, the unique characteristics that distinguish midsize organizations from enterprises, including risk profiles, resource constraints, organizational structures, and decision-making processes, have not been adequately incorporated into existing AI adoption models \cite{neumann2024exploring}. Current frameworks typically treat organizational size as a continuous variable rather than recognizing qualitative differences in adoption dynamics across organizational types \cite{jalil2025influential}.

Finally, there is a lack of validated, practical frameworks that provide actionable guidance specifically designed to accommodate both resource constrained midsize organizations and complexity constrained enterprises \cite{hussain2024strategic}. These gaps necessitate the development of a specialized framework that addresses the distinctive needs, constraints, and opportunities of organizations across these scales in their GenAI adoption journey. The FAIGMOE framework, presented in subsequent sections, is designed to fill these critical gaps through a tailored, scalable, evidence-based approach to GenAI implementation.

\section{Theoretical Foundation}
\label{sec:methodology}

The FAIGMOE framework is anchored in a comprehensive multi-theoretical\footnote{A Multi-Theory Framework (MTF) combines various theoretical perspectives to offer a deeper and more holistic understanding of complex phenomena.} foundation that synthesizes established models of technology adoption, organizational behavior, and innovation management. This theoretical integration provides the conceptual architecture necessary to understand the complex interplay of drivers, constraints, and enablers that influence GenAI adoption across diverse organizational contexts, from resource constrained midsize organizations to complexity constrained larger enterprises.

\subsection{Core Theoretical Underpinnings of FAIGMOE}

The development of FAIGMOE is grounded in three primary theoretical perspectives that collectively address the multifaceted nature of GenAI adoption across organizational scales:

\textbf{Technology-Organization-Environment (TOE) Framework:} Originally developed by Tornatzky and Fleischer \cite{baker2011technology}, the TOE framework provides a holistic lens for examining technology adoption by considering three critical contexts: technological readiness and characteristics, organizational capabilities and structure, and environmental pressures and opportunities \cite{sanchez2025artificial}. Within FAIGMOE, the TOE framework serves as the primary organizing structure, informing how technological capabilities, organizational resources, and external market forces interact to influence GenAI adoption decisions across different organizational scales \cite{donmez2025understanding, bhuiyan2024industry}. The framework recognizes that technological readiness manifests differently in midsize organizations (focusing on cloud-based solutions and vendor partnerships) versus enterprises (emphasizing system integration and legacy infrastructure compatibility).

\textbf{Technology Acceptance Model (TAM):} While Davis's original TAM \cite{davis1989technology} focused on individual-level technology acceptance, FAIGMOE adapts and extends these constructs to the organizational level across different scales. The framework incorporates perceived usefulness and perceived ease of use as critical determinants of organizational GenAI adoption, while recognizing that these perceptions are mediated by organizational characteristics such as leadership attitudes, resource availability, technical expertise, and organizational complexity \cite{dalle2024cultural, hadian2024technology}. In midsize organizations, perceived ease of use carries greater weight due to limited technical resources, while enterprises prioritize perceived usefulness across multiple business units \cite{guevara2024enhancing}.

\textbf{Diffusion of Innovations (DOI) Theory:} Rogers's DOI theory \cite{rogers2014diffusion} contributes essential insights into how GenAI innovations spread within and across organizations of different scales. FAIGMOE leverages DOI's innovation characteristics relative advantage, compatibility, complexity, trialability, and observability to assess GenAI solutions and design implementation strategies that enhance adoption likelihood \cite{costa2025unveiling}. The framework particularly emphasizes trialability and observability as critical factors for risk-averse midsize organizations, while addressing compatibility and complexity concerns more prominent in enterprises with established processes and multiple stakeholder groups \cite{yosua2019opinion}.

These theoretical foundations collectively enable FAIGMOE to address not merely adoption decisions, but the entire lifecycle of GenAI integration, institutionalization, and sustained value realization across organizational contexts ranging from agile midsize firms to complex enterprise environments.

\subsection{Integration of Organizational Change and Management Theories}

Recognizing that GenAI adoption represents a significant organizational transformation with scale dependent dynamics, FAIGMOE integrates established theories of organizational change and strategic management to address the human and structural dimensions of technology implementation:

\textbf{Kotter's 8-Step Change Model:} Kotter's systematic approach to organizational change \cite{kotter2012heart} provides the methodological foundation for FAIGMOE's change management components. The framework incorporates Kotter's emphasis on creating urgency, building guiding coalitions, developing clear vision and strategy, communicating transformation vision, empowering broad-based action, generating short-term wins, sustaining acceleration, and institutionalizing new approaches \cite{waren2025potential}. Implementation of these steps differs between midsize organizations (with shorter communication chains and more direct leadership engagement) and enterprises (requiring formal change management structures and multi-level stakeholder coordination) \cite{heikkila2024implementing}.

\textbf{McKinsey 7S Framework:} The 7S model's systems perspective \cite{singh2013study} informs FAIGMOE's holistic approach to organizational alignment. The framework ensures that GenAI implementation considers the interdependencies between strategy, structure, systems, shared values, skills, staff, and style, preventing the common pitfall of treating technology adoption as purely a technical exercise \cite{abouaomar2024overcoming}. The complexity of achieving alignment across these seven dimensions increases with organizational scale, requiring different coordination mechanisms in midsize organizations versus enterprises \cite{vinayavekhin2023roadmapping}.

\textbf{Resource-Based View (RBV):} Barney's RBV \cite{barney1991firm} emphasizes the strategic importance of unique organizational resources and capabilities in creating sustainable competitive advantage. FAIGMOE applies RBV principles to help organizations across scales identify, develop, and leverage AI-relevant resources including data assets, technical capabilities, human capital, and organizational learning capacity \cite{ashfaq2025strategic}. Midsize organizations typically focus on building partnerships and leveraging external resources, while enterprises emphasize developing internal capabilities and optimizing existing resource portfolios \cite{willie2025leveraging, kaur2024resource}.

\textbf{Absorptive Capacity Theory:} Cohen and Levinthal's concept of absorptive capacity \cite{cohen1990absorptive}—the ability to recognize, assimilate, and apply new knowledge is particularly relevant for GenAI adoption across organizational scales. FAIGMOE incorporates absorptive capacity assessment and development as critical components of organizational readiness and capability building \cite{abou2023impact}. The framework recognizes that absorptive capacity challenges differ between midsize organizations (limited specialist expertise) and enterprises (knowledge distribution across units and coordination challenges) \cite{sancho2022impact}.

\subsection{Conceptual Model Development}

The synthesis of these theoretical foundations culminates in FAIGMOE's conceptual architecture: a staged, modular framework comprising four interconnected phases that reflect both linear progression and iterative refinement, with scalable components addressing diverse organizational contexts:

\textbf{Phase 1 - Strategic Assessment:} This foundational phase evaluates organizational readiness across multiple dimensions including digital maturity, resource availability, cultural alignment, and strategic fit for GenAI adoption \cite{van2024bridging}. Drawing from TOE framework constructs, this phase systematically assesses technological infrastructure, organizational capabilities, and environmental factors that influence adoption success across different organizational scales \cite{prakash2025evaluating}.

\textbf{Phase 2 - Adoption Planning:} Building on assessment outcomes, this phase focuses on strategic use case identification, comprehensive risk assessment, governance framework development, and change management strategy formulation \cite{gohil2025managing}. TAM constructs inform the evaluation of perceived usefulness and implementation complexity for prioritized use cases, with planning approaches adapted to organizational scale and complexity \cite{ishengoma2024revisiting}.

\textbf{Phase 3 - Implementation and Integration:} This phase encompasses technical deployment, business process reengineering, workforce development, and stakeholder engagement activities \cite{popoola2024cross}. DOI theory guides the design of implementation approaches that maximize trialability and observability while addressing scale-specific challenges such as resource constraints in midsize organizations and coordination complexity in enterprises \cite{anand2024theory}.

\textbf{Phase 4 - Operationalization and Optimization:} The final phase focuses on embedding GenAI capabilities into standard operating procedures, establishing performance measurement systems, and developing scaling strategies for broader organizational deployment \cite{smith2025strategic}. Change management theories inform the institutionalization of new practices and continuous improvement processes, with approaches tailored to organizational structure and decision-making processes \cite{monferdini2024businesses}.

Each phase incorporates feedback loops and iteration mechanisms, recognizing that GenAI adoption is not a linear process but rather an adaptive journey requiring continuous adjustment based on learning and changing circumstances across organizational contexts \cite{rahouli2025generative}.

\subsection{Framework Assumptions and Scope}

FAIGMOE operates within a defined scope and set of assumptions that establish its applicability across organizational scales:

\textbf{Organizational Scope:} The framework targets both midsize organizations (typically 50-250 employees, \$10M-\$1B annual revenue) and larger enterprises (1,000+ employees, \$1B+ revenue) \cite{dorr2024state}. Midsize organizations are characterized by operational flexibility, flatter structures, and resource constraints, while enterprises exhibit greater complexity, more formal processes, and extensive resources but face coordination challenges \cite{dewi2025structure}.

\textbf{Technology Scope:} FAIGMOE adopts a technology-agnostic approach\footnote{A technology-agnostic approach is a flexible strategy that focuses on selecting the most effective tools and technologies for a given problem, without being restricted to any particular platform, framework, or vendor.}, accommodating various GenAI platforms and applications including large language models, multimodal AI systems, code generation tools, and creative AI platforms. The framework emphasizes commercially available solutions suitable for both midsize organizations (cloud-based, managed services) and enterprises (hybrid deployments, custom integrations).

\textbf{Implementation Scope:} The framework focuses on internal organizational adoption and integration processes rather than external product development or commercialization of GenAI capabilities. It addresses operational efficiency, decision support, and process automation applications across organizational functions, acknowledging scale-specific implementation patterns.

\textbf{Baseline Assumptions:} FAIGMOE assumes organizations possess fundamental digital literacy, basic IT infrastructure, and prior experience with enterprise software adoption. However, the framework recognizes that baseline capabilities vary significantly between midsize organizations and enterprises, with corresponding adaptations in implementation approaches.

\subsection{Methodological Principles Underlying FAIGMOE}

Beyond its theoretical foundations, FAIGMOE is guided by key methodological principles that ensure practical applicability and implementation success across organizational scales:

\textbf{Strategic Business Alignment:} All GenAI initiatives must demonstrate clear alignment with organizational strategy and measurable business value. This principle prevents technology-driven adoption in favor of business driven implementation, applicable across organizational scales with scale appropriate metrics and governance.

\textbf{Governance First Approach:} Proactive governance frameworks addressing ethics, risk management, compliance, and accountability are established before widespread deployment. This principle builds organizational trust and ensures responsible AI usage, with governance structures adapted to organizational complexity and resource availability.

\textbf{Cross Functional Orchestration:} Successful implementation requires coordinated effort across leadership, IT, legal, human resources, operations, and change management functions. This principle prevents siloed implementation and ensures organizational alignment, with coordination mechanisms adapted to organizational structure from informal collaboration in midsize firms to formal program management in enterprises.

\textbf{Modular Technical Architecture:} The framework supports integration with existing systems through modular approaches, APIs, and orchestration platforms that minimize disruption while maximizing interoperability\footnote{The capability of a system to operate with or utilize components or equipment from another system.}. Architectural approaches scale from simpler integrations suitable for midsize organizations to complex enterprise architectures.

\textbf{Agile and Iterative Scaling:} FAIGMOE advocates for pilot-driven implementation, rapid prototyping, continuous learning, and evidence based scaling decisions. This principle reduces implementation risk while accelerating time-to-value across organizational contexts, with scaling strategies adapted to organizational capacity and complexity.

These methodological principles serve as operational guidelines that bridge theoretical concepts with practical implementation, ensuring that FAIGMOE remains both academically rigorous and practically relevant for organizations across scales pursuing GenAI adoption \cite{sadek2025challenges}.

\section{The FAIGMOE Framework}
\label{sec:framework}

Building upon the theoretical foundations established in the previous section, this section presents the complete FAIGMOE framework: a structured, evidence-based methodology designed for GenAI adoption across midsize organizations and larger enterprises. The framework translates theoretical constructs into actionable implementation guidance, providing organizations with a systematic approach to navigate the complex journey from initial assessment through sustained optimization, with scalable components addressing the distinct needs of different organizational contexts.

\subsection{Framework Architecture and Core Components}

FAIGMOE employs a multi-dimensional architecture that recognizes the interconnected nature of technology adoption across organizational contexts. The framework integrates four critical dimensions that must be simultaneously addressed for successful GenAI implementation, with each dimension adapted to organizational scale and complexity:

\textbf{Strategic Dimension:} Ensures comprehensive alignment between GenAI initiatives and organizational strategic objectives, competitive positioning, and long-term value creation. This dimension incorporates strategic planning methodologies and portfolio management approaches adapted to organizational context from focused initiatives in midsize organizations to enterprise wide strategic programs in larger organizations.

\textbf{Operational Dimension:} Addresses workflow integration, process reengineering, governance mechanisms, and operational efficiency optimization. This dimension draws from business process management and operational excellence frameworks, recognizing that midsize organizations benefit from agile process adaptation while enterprises require formal process governance and cross unit coordination.

\textbf{Technical Dimension:} Encompasses infrastructure requirements, architectural design patterns, model lifecycle management, integration frameworks, and data ecosystem readiness. This dimension leverages enterprise architecture and systems integration best practices, scaling from cloud-native solutions appropriate for midsize organizations to hybrid architectures addressing enterprise legacy systems and complex integration requirements.

\textbf{Cultural Dimension:} Focuses on organizational change management, employee engagement, digital literacy development, ethical AI awareness, and resistance mitigation strategies. This dimension applies change management and organizational behavior principles, recognizing that midsize organizations leverage informal communication and direct leadership engagement while enterprises require formal change management structures and multi-level stakeholder coordination.

The framework's modular design enables organizations across scales to customize implementation approaches based on their specific context, resources, structural complexity, and strategic priorities while maintaining methodological consistency. Component dependencies are explicitly mapped to ensure proper sequencing and resource allocation throughout the implementation process, with complexity adjusted to organizational capacity.

\begin{figure}[htbp]
\centering
\begin{tikzpicture}[node distance=1.6cm, every node/.style={align=center}]
    \node (assessment) [rectangle, draw, fill=blue!10, rounded corners, minimum width=5cm] 
        {Phase 1: Strategic Assessment \\ \footnotesize Readiness, Capabilities, Risk Analysis};
    \node (planning) [rectangle, draw, fill=green!10, below of=assessment, rounded corners, minimum width=5cm] 
        {Phase 2: Planning \& Use Case Development \\ \footnotesize Strategic Alignment, Prioritization, Roadmaps};
    \node (implementation) [rectangle, draw, fill=orange!10, below of=planning, rounded corners, minimum width=5cm] 
        {Phase 3: Implementation \& Integration \\ \footnotesize Pilots, Scaling, Change Management};
    \node (optimization) [rectangle, draw, fill=purple!10, below of=implementation, rounded corners, minimum width=5cm] 
        {Phase 4: Operationalization \& Optimization \\ \footnotesize Monitoring, Continuous Improvement, Scaling};
    
    \draw[->, thick] (assessment) -- (planning);
    \draw[->, thick] (planning) -- (implementation);
    \draw[->, thick] (implementation) -- (optimization);
    
    \draw[->, thick, dashed, red] (optimization.east) -- ++(1.2,0) |- (assessment.east);
    \node[right, font=\tiny, text=red] at (5,-2) {Continuous\\Feedback\\Loop};
    
    \node[above of=assessment, node distance=1.2cm, font=\bfseries] {FAIGMOE Framework Architecture};
\end{tikzpicture}
\caption{Phased architecture of the FAIGMOE framework for GenAI adoption in midsize organizations and enterprises, showing both sequential progression and iterative feedback loops.}
\label{fig:faigmoe_phases}
\end{figure}
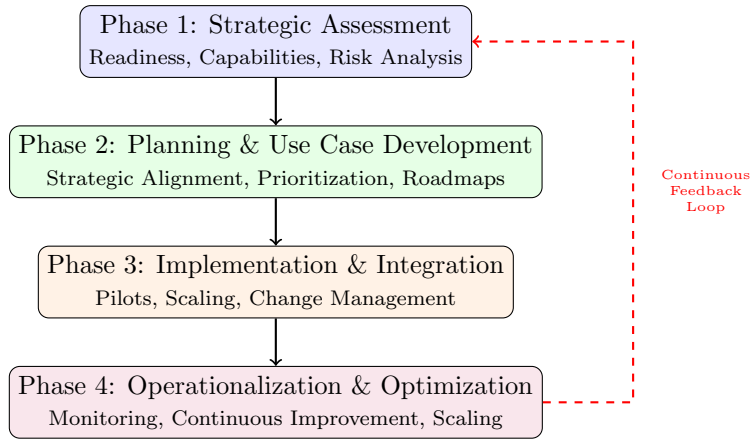

The framework's implementation follows a four phase structure as illustrated in Figure~\ref{fig:faigmoe_phases}, which depicts both the sequential progression through phases and the iterative feedback loops essential for adaptive implementation.

Each phase incorporates feedback loops and iteration mechanisms, recognizing that GenAI adoption is not a linear process but rather an adaptive journey requiring continuous adjustment based on learning and changing circumstances across organizational contexts \cite{hernandez2024adoption}.

\subsection{Phase 1: Strategic Assessment and Readiness Evaluation}

The foundational phase establishes a comprehensive baseline of organizational preparedness across multiple evaluation criteria, with assessment depth and breadth scaled to organizational context:

\textbf{Multi-Dimensional Readiness Evaluation:} A structured assessment methodology evaluates organizational maturity across five key areas: (1) strategic alignment and leadership commitment, (2) technical infrastructure and digital capabilities, (3) data quality, availability, and governance maturity, (4) organizational culture and change readiness, and (5) financial resources and investment capacity. Assessment complexity scales from streamlined evaluations for midsize organizations to comprehensive multi-unit assessments for enterprises.

\textbf{Capability Gap Analysis:} Systematic evaluation of existing human capital, technical competencies, and organizational processes identifies specific capability gaps that must be addressed for successful GenAI implementation. For midsize organizations, this typically emphasizes building external partnerships and selective capability development, while enterprises focus on internal capability optimization and knowledge distribution across business units.

\textbf{Comprehensive Risk Assessment:} A structured risk analysis methodology identifies and evaluates potential barriers including regulatory compliance challenges, data privacy concerns, ethical considerations, technical risks, and organizational resistance factors. Special attention is given to GenAI specific risks such as model hallucinations, bias amplification, intellectual property concerns, and content authenticity issues, with risk tolerance and mitigation strategies adapted to organizational scale and regulatory context.

\textbf{Deliverables:} The assessment phase produces a quantitative readiness scorecard, detailed capability gap analysis, prioritized risk register with mitigation strategies, and evidence-based recommendations for proceeding with GenAI adoption. Deliverable detail and formality scale with organizational size and governance requirements.

\subsection{Phase 2: Strategic Planning and Use Case Development}

This phase transforms assessment insights into concrete implementation strategies and actionable project portfolios, with planning sophistication matched to organizational complexity:

\textbf{Strategic Alignment Framework:} Facilitated stakeholder engagement processes ensure GenAI initiatives directly support organizational strategic objectives. This includes development of AI vision statements, success metrics definition, and governance structure establishment. Midsize organizations typically employ streamlined alignment processes with direct executive involvement, while enterprises require formal program governance and multi stakeholder consensus building mechanisms.

\textbf{Use Case Identification and Prioritization:} A systematic methodology evaluates potential GenAI applications across organizational functions using multi-criteria decision analysis. Evaluation criteria include business value potential, technical feasibility, implementation complexity, resource requirements, risk level, and strategic fit. Priority use cases for midsize organizations typically focus on high-impact departmental applications, while enterprises prioritize cross-functional applications with enterprise wide scaling potential.

\textbf{Implementation Roadmap Development:} Detailed project planning creates phased implementation timelines with defined milestones, resource requirements, success criteria, and dependency mapping. The roadmap emphasizes pilot-first approaches that enable learning and risk mitigation before broader deployment, with scaling strategies adapted to organizational structure—from departmental rollouts in midsize firms to coordinated multi-unit deployments in enterprises.

\textbf{Governance Framework Design:} Comprehensive governance structures address ethical AI principles, risk management protocols, compliance requirements, performance monitoring systems, and accountability mechanisms. Governance complexity scales from streamlined oversight committees in midsize organizations to formal AI governance boards with established reporting structures in enterprises.

\subsection{Phase 3: Implementation and Integration}

The implementation phase executes strategic plans through systematic deployment of GenAI capabilities, with implementation approaches adapted to organizational capacity and complexity:

\textbf{Pilot Program Development:} Initial implementations focus on carefully selected, low risk use cases that demonstrate clear business value. Pilot designs incorporate comprehensive testing protocols, user feedback mechanisms, and performance measurement systems. Common pilot applications include intelligent document processing, automated customer inquiry responses, and internal knowledge management systems. Midsize organizations typically run focused pilots within single departments, while enterprises conduct parallel pilots across multiple business units to assess scalability.

\textbf{Technical Infrastructure Deployment:} Implementation of required technical components including cloud platforms, API management systems, security frameworks, monitoring tools, and integration middleware\footnote{Middleware is software that serves as a bridge between different applications, systems, and databases, offering shared services such as communication, data management, and security to ensure they operate together smoothly.}. Architecture decisions prioritize scalability, security, and maintainability while minimizing complexity and cost. Midsize organizations typically leverage cloud-native platforms and managed services, while enterprises balance cloud adoption with existing infrastructure integration and regulatory requirements.

\textbf{Organizational Change Implementation:} Systematic change management activities including stakeholder communication, training program delivery, support system establishment, and resistance management. Training programs focus on AI literacy, prompt engineering skills, responsible usage practices, and workflow integration techniques. Change approaches scale from informal communication and direct training in midsize organizations to formal change management programs with dedicated resources in enterprises.

\textbf{Cross Functional Team Coordination:} Implementation teams comprising business leaders, technical specialists, compliance officers, and change management professionals coordinate activities using agile methodologies and established project management frameworks. Team structures scale from core implementation teams in midsize organizations to program management offices coordinating multiple workstreams in enterprises.

\subsection{Phase 4: Operationalization and Continuous Optimization}

The final phase establishes sustainable operations and continuous improvement processes, with operational structures adapted to organizational scale:

\textbf{Performance Monitoring and Analytics:} Comprehensive monitoring systems track key performance indicators including system performance, user adoption rates, business impact metrics, and risk indicators. Advanced analytics identify optimization opportunities and emerging issues requiring attention. Monitoring sophistication scales from dashboard-based tracking in midsize organizations to enterprise analytics platforms with multi-dimensional reporting.

\textbf{Continuous Improvement Processes:} Systematic improvement methodologies incorporate user feedback, performance data, and emerging best practices to enhance system effectiveness and user experience. This includes model performance optimization, prompt engineering refinement, and workflow process enhancement. Improvement processes scale from agile iteration teams in midsize organizations to formalized continuous improvement programs in enterprises.

\textbf{Knowledge Management and Scaling:} Organizational learning processes capture implementation experiences, best practices, and lessons learned for broader organizational benefit. Centers of Excellence (CoEs)\footnote{A Center of Excellence (CoE) is a dedicated group of specialists that delivers leadership, best practices, guidance, and expertise in a particular domain, such as a technology, process, or application, to enhance organizational efficiency and encourage broader adoption.}or similar structures facilitate knowledge sharing and support scaling to additional use cases and organizational units. Knowledge management approaches range from informal communities of practice in midsize organizations to formal CoEs with dedicated resources in enterprises.

\textbf{Strategic Evolution:} Regular strategic reviews assess GenAI portfolio performance, identify new opportunities, and adjust strategies based on evolving business needs and technological capabilities. Review frequency and formality scale with organizational complexity, from quarterly executive reviews in midsize firms to formal portfolio governance processes in enterprises.

\subsection{Framework Implementation Success Factors}

Successful FAIGMOE implementation requires attention to several critical success factors derived from empirical research and best practice analysis across organizational scales:

\textbf{Executive Leadership and Sponsorship:} Strong, visible leadership commitment provides necessary resources, removes organizational barriers, and demonstrates strategic importance. Leadership engagement manifests differently across scales from direct CEO involvement in midsize organizations to C-suite sponsorship\footnote{C-suite sponsorship involves a senior executive at the C-level providing strategic support, advocacy, and resource allocation to advance the success of a specific individual, project, or initiative.} and delegated program leadership in enterprises.

\textbf{Stakeholder Engagement and Communication:} Comprehensive stakeholder engagement strategies ensure broad organizational buy-in and address concerns proactively. Communication approaches scale from direct engagement in midsize organizations to structured communication programs in enterprises.

\textbf{Iterative Implementation Approach:} Phased implementation with regular evaluation and adjustment opportunities reduces risk while enabling organizational learning. Iteration cycles adapt to organizational decision-making speed—faster in midsize organizations, more deliberate in enterprises.

\textbf{Capability Development Investment:} Sustained investment in human capital development, technical infrastructure, and organizational processes supports long-term success. Investment strategies differ between midsize organizations (targeted, partnership focused) and enterprises (comprehensive, internally-focused).

\textbf{Governance and Risk Management:} Robust governance frameworks and proactive risk management build organizational confidence and ensure responsible AI deployment. Governance formality and structure scale with organizational complexity and regulatory requirements.

These success factors provide practical guidance for organizations across scales implementing FAIGMOE while highlighting common pitfalls that must be avoided to ensure sustainable GenAI adoption and integration.

\section{Framework Validation and Application}
\label{sec:validation}

To establish the credibility, applicability, and practical utility of the FAIGMOE framework across diverse organizational contexts, a validation strategy was designed employing multiple research methodologies. This section presents the proposed validation methodology and demonstrates the framework's conceptual applicability through illustrative scenarios representing both midsize organizations and enterprises.

\subsection{Proposed Validation Methodology}

A rigorous mixed methods research design is proposed to validate FAIGMOE across organizational scales \cite{tang2025using, lopez2023importance}:

\textbf{Methodological Triangulation:} The validation strategy should incorporate three complementary research approaches \cite{shapland2024quantitative}:

\textit{Documentary Evidence Analysis:} Systematic review of organizational strategic documents, digital transformation roadmaps, and AI readiness assessments from both midsize organizations and enterprises across diverse industries to benchmark framework assumptions against real-world contexts \cite{aldoseri2024methodological}.

\textit{Stakeholder Interview Protocol:} Semi-structured interviews with key stakeholders including executives, technology leaders, and implementation teams from organizations across scales to explore GenAI adoption practices, challenges, and framework relevance \cite{kongsten2024frameworks}.

\textit{Longitudinal Implementation Studies:} Direct application and observation of framework components in live organizational environments to provide empirical evidence of practical feasibility and effectiveness \cite{jacob2025ai}.

\textbf{Validation Criteria:} The validation process should evaluate FAIGMOE against established criteria:
- \textit{Theoretical Rigor:} Consistency with established theories and conceptual coherence
- \textit{Practical Utility:} Applicability and usefulness in real organizational contexts
- \textit{Contextual Relevance:} Appropriateness for both midsize and enterprise characteristics
- \textit{Implementation Feasibility:} Realistic resource requirements across organizational scales
- \textit{Scalability and Adaptability:} Flexibility across different organizational contexts

\subsection{Illustrative Application Scenarios}

To demonstrate FAIGMOE's conceptual applicability, we present illustrative scenarios representing typical GenAI adoption challenges across organizational scales:

\textbf{Scenario 1: Midsize Financial Services Organization}

A regional financial institution seeks to implement GenAI for customer service automation while managing strict regulatory requirements. The organization faces resource constraints and limited in-house AI expertise, which requires external partnerships and focused implementation approaches. FAIGMOE's Assessment phase would identify compliance requirements and capability gaps, while the Planning phase would prioritize low-risk applications aligned with regulatory frameworks. The modular implementation approach enables focused deployment without overwhelming organizational capacity.

\textbf{Scenario 2: Midsize Healthcare Technology Firm}

A health technology company developing clinical software aims to leverage GenAI for internal research processes. The organization requires high accuracy standards and explainability given healthcare applications. FAIGMOE's framework addresses these requirements through comprehensive risk assessment, governance framework development emphasizing transparency, and structured change management ensuring clinical staff understand AI limitations and appropriate usage.

\textbf{Scenario 3: Enterprise Retail Organization}

A national retail enterprise with multiple business units and geographic locations seeks enterprise wide GenAI adoption for supply chain optimization and customer experience. The organization faces coordination challenges across units and requires formal governance structures. FAIGMOE's scalable architecture supports coordinated implementation across divisions while maintaining consistency, with enterprise governance mechanisms addressing multi-stakeholder complexity.

\textbf{Scenario 4: Professional Services Enterprise}

A large consulting firm with multiple practice areas requires GenAI capabilities while maintaining client confidentiality and professional standards. The organization benefits from centralized infrastructure but needs practice-level customization. FAIGMOE accommodates this through modular implementation enabling practice-specific applications within enterprise governance frameworks, balancing standardization with necessary flexibility.

\subsection{Expert Validation Approach}

A comprehensive expert validation process using modified Delphi methodology\footnote{The Delphi method is a structured communication approach designed to reach a consensus among a group of experts by using multiple rounds of anonymous surveys.} is proposed to assess framework validity \cite{villarino2024conceptualization}:

\textbf{Proposed Panel Composition:} An expert panel should include \cite{kellerhuis2025expert}:
- Academic researchers specializing in information systems, technology adoption, and AI governance
- Senior practitioners from both midsize organizations and enterprises across industries
- Implementation consultants with experience across organizational scales

\textbf{Validation Protocol:} A three-round structured process should evaluate \cite{pires2024guidelines}:
- Theoretical foundations and conceptual coherence
- Practical applicability across organizational contexts
- Implementation feasibility and resource requirements
- Framework refinements based on expert feedback

\subsection{Future Empirical Validation Requirements}

While the framework is theoretically grounded and conceptually validated, comprehensive empirical validation remains essential. Future research should pursue:

\textbf{Longitudinal Implementation Studies:} Multi-year studies tracking organizations through complete FAIGMOE implementation cycles across both midsize and enterprise contexts.

\textbf{Comparative Effectiveness Research:} Rigorous comparison of FAIGMOE-guided implementations versus alternative approaches to establish relative effectiveness.

\textbf{Large-Scale Survey Research:} Quantitative assessment of framework adoption patterns, success factors, and organizational outcomes across diverse contexts.

\textbf{Industry Specific Validation:} Detailed validation within regulated industries requiring specialized compliance protocols.

These future validation efforts will strengthen empirical foundations and enable data-driven framework refinements as GenAI technologies and organizational practices evolve.

\section{Discussion}
\label{sec:discussion}

The development of the FAIGMOE framework represents a significant contribution to the growing body of knowledge on organizational AI adoption, specifically addressing both midsize organizations and larger enterprises that face distinct yet underserved adoption challenges. This section provides critical analysis of the framework's distinctive contributions, its effectiveness in addressing scale-specific organizational challenges, scalability considerations, and acknowledges inherent limitations that bound its applicability.

\subsection{Theoretical and Practical Contributions to AI Adoption Literature}

FAIGMOE makes several significant contributions that distinguish it from existing technology adoption frameworks and AI implementation methodologies. Table~\ref{tab:framework_comparison} illustrates how FAIGMOE addresses gaps in existing frameworks through its combination of GenAI specific guidance, multi-scale applicability, and actionable implementation protocols.

\begin{table}[htbp]
\centering
\caption{Comparative Analysis of Technology Adoption Frameworks}
\label{tab:framework_comparison}
\begin{tabular}{p{2cm} p{3cm} p{2.5cm} p{1.5cm} p{3cm}}
\toprule
\textbf{Framework} & \textbf{Theoretical Focus} & \textbf{Organizational Scale} & \textbf{GenAI Specificity} & \textbf{Implementation Guidance} \\
\midrule
\textbf{TAM} & Individual acceptance & Generic & Low & Conceptual only \cite{davis1989technology} \\
\midrule
\textbf{TOE} & Multi-level factors & Generic & Low & Conceptual only \cite{prakash2025evaluating} \\
\midrule
\textbf{DOI} & Innovation diffusion & Generic & Low & Conceptual only \cite{rogers2014diffusion} \\
\midrule
\textbf{AI Maturity Models} & Capability levels & Large enterprises & Medium & Limited \cite{sadiq2021artificial} \\
\midrule
\textbf{FAIGMOE (our framework)} & Multi-theoretical integration & Midsize \& Enterprise & High & Comprehensive protocols \\
\bottomrule
\end{tabular}
\end{table}

\textbf{Integration of Multi-Theoretical Perspectives:} Unlike traditional frameworks that rely on single theoretical lenses, FAIGMOE synthesizes insights from technology adoption theory (TOE, TAM, DOI), organizational change management, and innovation management to create a more comprehensive understanding of GenAI implementation dynamics across organizational scales. This theoretical integration addresses the complexity of GenAI adoption that cannot be adequately explained by any single theoretical framework, particularly when considering the different manifestations of adoption challenges in resource-constrained versus complexity-constrained environments.

\textbf{GenAI Specific Implementation Guidance:} Existing technology adoption models provide general frameworks but lack specificity for generative AI technologies. FAIGMOE addresses GenAI-specific challenges including prompt engineering, model hallucination management, retrieval-augmented generation implementation, ethical AI considerations, and content authenticity verification. This specificity transforms abstract adoption concepts into actionable implementation protocols adaptable to different organizational contexts and resource profiles.

\textbf{Operational Prescriptiveness Across Scales:} While traditional frameworks excel at explaining adoption factors, they typically provide limited operational guidance for implementation \cite{horani2025critical}. FAIGMOE bridges this theory-practice gap by offering detailed implementation protocols, assessment instruments, governance templates, and change management procedures that organizations can adapt to their specific contexts. Importantly, the framework provides scale-appropriate guidance, recognizing that midsize organizations require streamlined approaches while enterprises need comprehensive coordination mechanisms.

\textbf{Contextual Adaptation for Organizational Diversity:} Most existing frameworks assume either resource abundant large enterprise contexts or very small business environments, leaving a significant gap for midsize organizations and many mid-tier enterprises \cite{qu2025artificial}. FAIGMOE specifically addresses both resource constraints (typical of midsize organizations) and complexity constraints (characteristic of enterprises), representing a significant advancement in context-specific framework development.

\subsection{FAIGMOE Effectiveness in Addressing Scale-Specific Organizational Challenges}

Our proposed framework demonstrates particular strength in addressing the distinctive challenges that characterize technology adoption across organizational scales:

\textbf{Resource Optimization and Complexity Management:} FAIGMOE's phased approach enables midsize organizations to optimize limited resources through strategic sequencing and component reuse, while helping enterprises manage implementation complexity through structured coordination mechanisms. The framework's emphasis on appropriate technology choices cloud native solutions for midsize organizations, hybrid architectures for enterprises enables effective implementation across contexts.

\textbf{Risk Mitigation Across Contexts:} The framework adopts a pilot first approach and a comprehensive risk assessment methodology to address different risk profiles: the lower risk tolerance of midsize organizations due to resource constraints, and the concerns of enterprises regarding large scale implementation failures and reputational risks. By emphasizing appropriate initial implementations for each context, FAIGMOE enables organizations to build confidence in a systematic manner.

\textbf{Capability Building at Scale:} FAIGMOE's integrated approach to capability development addresses different organizational needs: midsize organizations require focused skill development and external partnerships, while enterprises need knowledge distribution across units and coordination capability development. The framework's emphasis on developing appropriate organizational capabilities reduces both external dependence (for midsize organizations) and coordination failures (for enterprises).

\textbf{Scalable Governance Approaches:} The proposed framework provides governance structures that scale appropriately: streamlined oversight for midsize organizations without excessive bureaucracy, and formal governance frameworks for enterprises requiring accountability across multiple stakeholders. This balance ensures responsible AI implementation without creating inappropriate organizational burden.

\subsection{Scalability, Adaptability, and Generalizability}

FAIGMOE's design incorporates several features that enhance its applicability across diverse organizational contexts:

\textbf{Modular Architecture Benefits:} The proposed framework's modular design enables organizations across scales to select and customize components based on their specific needs, maturity levels, and resource availability. This modularity supports implementation approaches ranging from focused departmental deployments (midsize organizations) to coordinated enterprise wide programs (large enterprises).

\textbf{Technology Agnosticism:} By avoiding vendor specific recommendations, FAIGMOE maintains relevance across different technology platforms and enables organizations to select solutions matching their infrastructure, regulatory requirements, and strategic relationships  This approach provides resilience against rapid technological evolution while accommodating different procurement and vendor management approaches across organizational scales.

\textbf{Industry and Regulatory Adaptability:} Our framework demonstrates conceptual adaptability across industry contexts and regulatory environments while maintaining core methodological consistency. The modular structure enables customization for sector specific requirements without compromising implementation effectiveness.

\textbf{Organizational Maturity Accommodation:} FAIGMOE's assessment-driven approach enables adaptation to different levels of digital maturity and AI readiness, making it conceptually applicable to organizations across the digital transformation spectrum and organizational scales.

\subsection{Limitations and Boundary Conditions}

As a perspective framework, FAIGMOE's limitations must be acknowledged to define its appropriate application context and identify areas requiring further development:

\textbf{Empirical Validation Requirements:} While theoretically grounded and conceptually validated, FAIGMOE requires comprehensive empirical validation across diverse organizational contexts, industries, and implementation scenarios. Longitudinal studies tracking organizations through complete implementation cycles are essential to fully establish effectiveness and identify necessary refinements.

\textbf{Regulatory Environment Constraints:} While FAIGMOE incorporates governance considerations, organizations in highly regulated sectors may require additional specialized compliance protocols beyond the framework's current scope. Sector specific adaptations for healthcare, financial services, defense, and other heavily regulated industries represent important areas for future framework development.

\textbf{Digital Infrastructure Prerequisites:} FAIGMOE assumes baseline digital infrastructure maturity including cloud connectivity, data management capabilities, and cybersecurity frameworks. Organizations lacking these prerequisites may need foundational digital transformation before effectively applying the framework.

\textbf{Organizational Readiness Assumptions:} The framework assumes leadership commitment to AI adoption and basic change management capabilities. Organizations with significant cultural resistance or limited change management experience may require preparatory organizational development.

\textbf{Technological Evolution Challenges:} The rapid evolution of GenAI technologies necessitates ongoing framework updates to maintain relevance. As new capabilities emerge—including agentic AI, multimodal systems, and advanced reasoning models—the framework must evolve to address new implementation considerations and challenges.

\textbf{Scale Boundary Considerations:} While designed for midsize organizations and enterprises, the framework's optimal applicability range requires empirical validation. Very small organizations (under 50 employees) may find components overly complex, while very large multinational corporations may require additional coordination mechanisms.

\subsection{Implications for Research and Practice}

The development of FAIGMOE has several important implications for both academic research and organizational practice:

\textbf{Research Implications:} FAIGMOE demonstrates the value of developing scale-aware, context-specific technology adoption frameworks rather than relying solely on generic models. The framework's multi-theoretical foundation provides a template for future research on complex technology adoption phenomena across organizational scales. The proposed validation methodology offers a robust approach for future framework development research. Additionally, the framework highlights important research questions regarding how organizational scale influences technology adoption dynamics beyond simple resource availability.

\textbf{Practice Implications:} For organizations across scales, FAIGMOE provides a conceptual roadmap for GenAI adoption that balances theoretical rigor with practical applicability. For consultants and technology vendors, the framework offers a structured approach to supporting diverse organizational contexts. For policymakers, FAIGMOE highlights the importance of developing differentiated resources and support mechanisms recognizing that organizations face qualitatively different challenges based on scale and structure.

These implications suggest that FAIGMOE represents both a practical tool for organizations and a contribution to understanding how complex technologies can be successfully adopted across diverse organizational contexts with different resource profiles and structural characteristics.

\section{Practical Implications}

Beyond its theoretical contributions, the FAIGMOE framework provides conceptual guidance for practitioners navigating the complex landscape of GenAI adoption across midsize organizations and enterprises. This section translates framework concepts into operational guidance, offering implementation protocols, resource recommendations, and strategic considerations derived from theoretical analysis and best practice synthesis.

\subsection{Strategic Implementation Guidelines for Practitioners}

Effective FAIGMOE implementation requires systematic execution adapted to organizational scale and complexity. The following guidelines provide a structured approach for practitioners across organizational contexts:

\textbf{Phase 1: Foundation and Assessment}

\textit{Conduct Comprehensive Readiness Assessment:} Utilize multi-dimensional assessment instruments adapted to organizational scale to evaluate preparedness across technical infrastructure, data maturity, governance frameworks, cultural readiness, and leadership commitment \cite{tasleem2023organizational}. Midsize organizations should focus on identifying critical capability gaps and partnership opportunities, while enterprises should assess coordination capabilities and knowledge distribution mechanisms.

\textit{Secure Executive Sponsorship and Governance:} Establish executive sponsorship and governance structures appropriate to organizational context streamlined oversight committees for midsize organizations, formal governance boards for enterprises \cite{sayles2024ai}. Governance frameworks should address decision rights, accountability mechanisms, risk management protocols, and escalation procedures scaled to organizational complexity \cite{de2021artificial}.

\textit{Establish Baseline Metrics:} Define current-state performance metrics across targeted business processes to enable quantitative evaluation of GenAI impact \cite{ghafoori2025ai,blagec2020critical}. Metric sophistication should match organizational capabilities and reporting requirements.

\textbf{Phase 2: Strategic Planning and Prioritization}

\textit{Align GenAI Initiatives with Strategic Objectives:} Ensure proposed GenAI applications directly support organizational strategic priorities \cite{denni2024navigating}. Alignment processes should reflect organizational decision-making structures direct executive involvement for midsize organizations, formal program governance for enterprises \cite{smith2025strategic}.

\textit{Apply Rigorous Use Case Prioritization:} Employ structured prioritization frameworks evaluating potential applications across business value, technical feasibility, implementation complexity, resource requirements, risk profile, and strategic fit \cite{ostrowski2025using}. Midsize organizations should emphasize focused, high-impact departmental applications, while enterprises should consider cross-functional applications with scaling potential.

\textit{Develop Detailed Implementation Roadmaps:} Create phased implementation plans with defined milestones, resource requirements, and success criteria. Roadmap complexity should match organizational coordination capabilities streamlined plans for midsize organizations, comprehensive program roadmaps for enterprises.

\textbf{Phase 3: Implementation and Integration}

\textit{Adopt Pilot First Approaches:} Begin with carefully designed pilot implementations that enable controlled experimentation. Pilot scope should reflect organizational capacity—single-department pilots for midsize organizations, multi-unit parallel pilots for enterprises assessing scalability \cite{tahmasebinia2024piloting}.

\textit{Build Cross Functional Implementation Teams:} Assemble teams representing IT, data science, business operations, compliance, and change management \cite{haque2025algorithmic}. Team structures should scale from core implementation teams (midsize organizations) to program management offices coordinating multiple workstreams (enterprises).

\textit{Implement Agile Development Methodologies:} Apply agile principles including iterative development, continuous feedback, and incremental delivery \cite{mahboob2024future}. Agile practices should adapt to organizational decision-making speed and approval processes.

\textit{Embed Comprehensive Change Management:} Integrate change management activities throughout implementation \cite{ostrowski2025using}. Change approaches should scale from informal communication and direct training (midsize organizations) to formal change management programs with dedicated resources (enterprises).

\textbf{Phase 4: Operationalization and Scaling}

\textit{Establish Performance Monitoring Systems:} Implement monitoring frameworks appropriate to organizational sophistication dashboard based tracking for midsize organizations, enterprise analytics platforms for larger organizations \cite{valiulla2025empirical}.

\textit{Develop Scaling Strategies:} Design systematic approaches for scaling successful pilots \cite{hendricks2025pilot}. Scaling should reflect organizational structure departmental rollouts for midsize firms, coordinated multi-unit deployments for enterprises.

\textit{Build Continuous Improvement Processes:} Establish mechanisms for capturing lessons learned and implementing enhancements \cite{li2025genai}. Improvement processes should match organizational formality and resource availability.

\subsection{Technology Selection Considerations}

Effective GenAI implementation requires careful technology selection balancing capability, cost, complexity, and organizational fit \cite{csahin2024generative}. Key considerations include:

\textbf{Foundation Models and Platforms:} Organizations should evaluate options including OpenAI, Azure OpenAI, Google, Anthropic, and AWS based on cost structure, compliance features, vendor support, and model capabilities \cite{jay2023enterprise}. Midsize organizations typically benefit from managed API services, while enterprises may consider hybrid deployment models.

\textbf{Application Development Frameworks:} Selection should consider development velocity, community support, and integration capabilities \cite{nguyen2025guidelines}. Framework choices should align with existing technical capabilities and development practices.

\textbf{Infrastructure and Security:} Infrastructure decisions should balance cloud-native solutions (appropriate for midsize organizations) with hybrid architectures addressing enterprise legacy systems and regulatory requirements \cite{shrivastava2024solutions}.

\textbf{Build vs. Buy Decisions:} Organizations should generally leverage managed services and commercial platforms for foundational capabilities while focusing internal development on differentiated applications \cite{singh2024transforming,mccarthy5234136buy}. Decision criteria should reflect organizational technical capacity and strategic priorities.

\subsection{Critical Success Factors}

Several factors significantly influence GenAI adoption success across organizational contexts \cite{shao2025unveiling}:

\textbf{Strategic and Organizational Factors:}

\textit{Clear Business Value Articulation:} Implementations should begin with explicitly defined business objectives and quantifiable success metrics \cite{chatterjee2025maximising}. Value articulation should match organizational sophistication and stakeholder expectations.

\textit{Sustained Executive Leadership:} Active executive sponsorship providing resources and removing barriers significantly influences success \cite{shields2024transformative}. Leadership engagement manifests differently across scales—direct CEO involvement (midsize) versus C-suite sponsorship with delegated program leadership (enterprises) \cite{beyrer2025key}.

\textit{Realistic Expectations Management:} Setting appropriate expectations regarding GenAI capabilities, limitations, and timelines prevents disappointment \cite{rahouli2025generative}.

\textbf{Technical and Operational Factors:}

\textit{Data Quality Prioritization:} Investment in data quality, accessibility, and governance fundamentally influences implementation success \cite{pathak2025leveraging}. Data strategies should reflect organizational data maturity and resource availability.

\textit{Human-Centered Design:} Solutions augmenting human capabilities rather than replacing workers generate greater acceptance \cite{manresa2025humanizing}. Design approaches should involve end-users appropriately.

\textit{Modular Architecture:} Technical architectures emphasizing modularity and API-based integration enable flexibility \cite{chenosov2025development}. Architectural complexity should match organizational technical sophistication.

\textbf{Governance and Risk Management:}

\textit{Proactive Governance:} Establishing governance frameworks before widespread deployment builds trust \cite{gandhi2025approaches}. Governance formality should scale with organizational complexity and regulatory requirements.

\textit{Continuous Monitoring:} Systematic monitoring enables proactive issue identification \cite{huang2024navigating}. Monitoring sophistication should match organizational capabilities.

\subsection{Common Implementation Challenges}

GenAI implementations frequently encounter predictable challenges:

\textbf{Unclear Business Value:} Technology-driven initiatives lacking strategic alignment often fail to demonstrate ROI. Mitigation requires structured business case development and explicit strategic alignment \cite{smith2025strategic}.

\textbf{Inadequate Data Readiness:} Poor data quality undermines model performance. Organizations should assess and improve data quality before implementation \cite{huang2024genai}.

\textbf{Underestimated Change Management:} User resistance and low adoption result from insufficient change management. Comprehensive change strategies involving users early are essential \cite{vuorenheimo2025ai}.

\textbf{Insufficient Governance:} Lack of governance frameworks risks ethics incidents and compliance violations. Early governance establishment is critical \cite{gandhi2025approaches}.

\textbf{Premature Scaling:} Scaling before adequate validation risks system failures. Comprehensive pilot validation and phased rollouts are essential \cite{bhattarai2025scaling}.

\textbf{Scale-Specific Challenges:} Midsize organizations face resource constraints and expertise gaps, while enterprises encounter coordination complexity and bureaucratic barriers \cite{rajaram2024generative}. Mitigation strategies should address context-specific challenges.

\subsection{Organizational Capability Development}

Long-term GenAI success requires systematic capability development \cite{tuomimaa2025generative}:

\textbf{AI Literacy Programs:} Comprehensive training addressing AI fundamentals, prompt engineering, and ethical considerations across organizational levels \cite{tadimalla2025ai}. Program sophistication should match organizational needs and resources.

\textbf{Knowledge Sharing Structures:} Organizations should establish structures facilitating expertise sharing and standard development—informal communities of practice for midsize organizations, formal Centers of Excellence for enterprises \cite{zhang2025research}.

\textbf{Continuous Learning:} Internal communities enabling peer learning and collaborative problem-solving support sustained capability development \cite{elsayary2024integrating}.

These practical implications provide conceptual guidance for organizations implementing FAIGMOE across diverse contexts, translating theoretical concepts into operational considerations while acknowledging that specific implementation approaches should be tailored to organizational circumstances.

\section{Future Research Directions}

While the FAIGMOE framework represents a conceptual contribution to GenAI adoption literature addressing both midsize organizations and enterprises, it reveals numerous opportunities for empirical investigation and theoretical refinement. This section identifies critical research directions that would advance both theoretical understanding and practical application of GenAI adoption frameworks across organizational scales, addressing gaps revealed through framework development.

\subsection{Empirical Validation and Longitudinal Studies}

As a perspective framework, FAIGMOE requires comprehensive empirical validation across diverse organizational contexts:

\textbf{Multi-Year Implementation Studies:} Research should track both midsize organizations and enterprises through complete FAIGMOE implementation cycles spanning multiple years to assess framework effectiveness across all phases and organizational scales. These studies should examine how implementation approaches evolve differently in resource-constrained versus complexity-constrained environments, which framework components prove most valuable across contexts, and how organizational scale influences long-term outcomes.

\textbf{Sustained Impact Measurement:} Future research must develop and validate comprehensive measurement frameworks for assessing long-term GenAI impacts on organizational performance, innovation capacity, competitive positioning, and workforce transformation across different organizational scales. Particular attention should be given to distinguishing scale-specific impact patterns and establishing causal relationships between framework implementation and organizational outcomes.

\textbf{Comparative Effectiveness Studies:} Rigorous comparative research should evaluate FAIGMOE effectiveness relative to alternative adoption approaches across midsize organizations and enterprises. These studies would provide evidence regarding which framework components contribute most significantly to successful outcomes under different organizational scales and conditions.

\textbf{Cross Scale Analysis:} Research examining how framework effectiveness varies between midsize organizations and enterprises would provide insights into scale dependent success factors and necessary adaptations.

\subsection{Industry and Sector Specific Framework Adaptations}

While FAIGMOE is conceptually adaptable across industries and organizational scales, empirical validation of sector specific adaptations is essential:

\textbf{Highly Regulated Industry Adaptations:} Research should develop and validate specialized framework extensions for sectors with stringent regulatory requirements including healthcare, financial services, education, and government. These adaptations should address how regulatory requirements interact with organizational scale, for example, how midsize financial institutions versus large banking enterprises navigate compliance differently.

\textbf{Industry Scale Interaction Studies:} Investigation of how industry characteristics and organizational scale jointly influence GenAI adoption patterns would provide nuanced implementation guidance. For instance, how do adoption dynamics differ between midsize and enterprise healthcare organizations compared to midsize and enterprise manufacturing firms?

\textbf{Geographic and Cultural Variations:} Cross-cultural research examining how national cultures, regulatory environments, and regional characteristics influence GenAI adoption across organizational scales would enhance framework global applicability.

\textbf{Organizational Archetype Development:} Research should identify distinct organizational archetypes spanning midsize and enterprise segments, developing tailored implementation guidance based on characteristics beyond simple employee counts or revenue figures.

\subsection{Human-AI Collaboration and Workforce Transformation Dynamics}

GenAI fundamentally transforms knowledge work across organizational scales, necessitating deeper investigation of human factors:

\textbf{Scale Dependent Workforce Impact Studies:} Research should examine how workforce impacts differ between midsize organizations and enterprises, including variations in job transformation patterns, skill requirement evolution, and organizational learning mechanisms.

\textbf{Cognitive and Behavioral Response Studies:} Investigation of how workers across different organizational contexts cognitively process and behaviorally adapt to AI-augmented workflows. This should include examining whether organizational scale influences acceptance patterns and trust development.

\textbf{Skills Evolution and Development:} Longitudinal research tracking skill requirements evolution in AI-augmented environments across organizational scales would inform differentiated workforce development strategies.

\textbf{Organizational Learning Mechanisms:} Investigation of how organizations of different scales develop and retain GenAI related knowledge, examining whether learning mechanisms differ between flatter midsize structures and complex enterprise hierarchies.

\textbf{Equity and Inclusion Considerations:} Studies examining how GenAI adoption affects different stakeholder groups across organizational scales would inform more equitable implementation approaches.

\subsection{Framework Digitalization and Decision Support Systems}

Opportunities exist to transform FAIGMOE from a conceptual framework into technology-enabled implementation support tools:

\textbf{Intelligent Assessment Instruments:} Development of AI powered assessment tools that automatically evaluate organizational readiness with scale-appropriate depth and recommend context-specific development initiatives \cite{yuldashev2024development}.

\textbf{Scale Adaptive Recommendation Systems:} Investigation of ML approaches for providing contextualized implementation guidance that adapts to organizational scale, industry context, and implementation history.

\textbf{Collaborative Implementation Platforms:} Development of digital platforms facilitating implementation coordination appropriate to organizational complexity, from simple collaboration tools for midsize organizations to sophisticated program management platforms for enterprises.

A Framework\subsection{Ethical, Legal, and Societal Implications}

As GenAI adoption scales across organizations, research must address expanding ethical, legal, and societal considerations \cite{mehler2025influencing}:

\textbf{Scale Dependent Governance Research:} Investigation of how organizational scale influences governance requirements, ethical considerations, and risk management approaches \cite{demidenko2010ethics}. Research should examine whether midsize organizations and enterprises face qualitatively different ethical challenges or simply different resource contexts for addressing similar challenges.

\textbf{Bias Detection and Mitigation:} Development of systematic methodologies for identifying and mitigating bias appropriate to different organizational contexts \cite{schwartz2022towards}.

\textbf{Transparency and Accountability:} Research examining optimal approaches for achieving appropriate transparency levels across organizational contexts, recognizing that stakeholder expectations and regulatory requirements may vary with organizational scale \cite{larsson2020transparency}.

\textbf{Societal Impact Assessment:} Broader studies examining how GenAI adoption across different organizational scales collectively impacts employment patterns, economic structures, and social dynamics \cite{vesnic2020societal}.

\subsection{Framework Evolution for Emerging Technologies}

Rapid AI technological advancement necessitates ongoing framework evolution \cite{amiri2024artificial}:

\textbf{Agentic AI Integration:} As autonomous AI agents emerge, research should examine implications for organizational workflows across scales, including how coordination complexity in enterprises versus resource constraints in midsize organizations influence agentic AI adoption. Particular attention should be given to recurring agentic AI patterns, such as autonomous task orchestration, multi-agent collaboration, and human-agent interfacing that shape integration strategies and operational outcomes \cite{acharya2025agentic}.

\textbf{Multimodal AI Applications:} Research examining organizational applications of multimodal AI systems across different organizational contexts would expand framework applicability \cite{soenksen2022integrated}.

\textbf{Infrastructure Evolution:} Investigation of edge computing, distributed architectures, and other emerging infrastructure paradigms across organizational scales \cite{gill2025edge}.

\textbf{Open-Source Ecosystem Development:} Research tracking open-source LLM ecosystem evolution and examining differential implications for midsize organizations versus enterprises \cite{bostrom2018strategic}.

\subsection{Research Priorities and Synthesis}

Given resource constraints in academic research, strategic prioritization is essential:

\textbf{Highest Priority:} Empirical validation studies across organizational scales and longitudinal impact assessment that would establish framework effectiveness and identify necessary refinements.

\textbf{High Priority:} Industry specific adaptations and cross scale comparative research that would provide practical guidance while advancing theoretical understanding of scale dependent adoption dynamics.

\textbf{Medium Priority:} Framework digitalization and emerging technology integration that would enhance long-term relevance and usability.

This research agenda provides a comprehensive roadmap for transforming FAIGMOE from a conceptual perspective framework into an empirically validated, practically proven approach to GenAI adoption across diverse organizational contexts.

\section{Conclusion}
\label{sec:conclusion}

Generative AI represents a transformative inflection point in organizational digital transformation. Midsize organizations (50-250 employees, \$10M-\$1B revenue) face resource constraints and limited expertise, while enterprises (1,000+ employees, \$1B+ revenue) encounter organizational complexity and coordination challenges. Both require specialized frameworks addressing their distinct adoption challenges.

This paper introduces FAIGMOE—the Framework for the Adoption and Integration of Generative AI in Midsize Organizations and Enterprises. The framework makes a significant theoretical contribution by integrating multiple perspectives (TOE, TAM, DOI, and organizational change theories) into a comprehensive model addressing GenAI adoption complexities across organizational scales. This multi-theoretical integration provides more complete explanation of adoption dynamics than single-theory approaches while offering practical applicability through scalable components.

FAIGMOE's four phase structure Strategic Assessment, Planning and Use Case Development, Implementation and Integration, and Operationalization and Optimization translates theoretical constructs into actionable guidance. The modular design accommodates both resource constrained midsize organizations and complexity constrained enterprises while incorporating GenAI specific considerations including prompt engineering, model orchestration, and hallucination management.

As a perspective framework, FAIGMOE provides conceptual foundations requiring empirical validation through longitudinal studies, sector specific adaptations, and investigation of human-AI collaboration dynamics. The framework is designed as an evolving solution requiring continuous refinement as GenAI technologies mature. By providing theoretically grounded guidance across organizational scales, FAIGMOE contributes to more equitable and sustainable AI adoption, enabling organizations to participate effectively in AI-driven transformation while ensuring responsible implementation.


\bibliographystyle{IEEEtran}
\bibliography{ref.bib}

\end{document}